\documentstyle[amssymb,aps]{revtex}

\begin{document}
\title{The Graviton and the Nature of Dark Energy\thanks{%
*To appear in the Proceedings of the Cosmology Particle Astrophysics COSPA
conference, Taipei, November 2003.}}
\author{A. Zee}
\address{Kavli Institute for Theoretical Physics\\
University of California\\
Santa Barbara, CA 93106\\
USA\\
zee@kitp.ucsb.edu}

\begin{abstract}
I discuss various thoughts, old and new, about the cosmological constant (or
dark energy) paradox. In particular, I suggest the possibility that the
cosmological ``constant'' may decay as $\Lambda \sim \alpha
^{2}m_{N}^{3}/\tau ,$ where $\tau $ is the age of the universe.
\end{abstract}

\bigskip

\section{\protect\bigskip The paradox}

Physics thrives on paradoxes. The observation of dark energy,\cite{riess}%
\cite{perl} which may or may not be a cosmological constant, has spread
despondency and elation around the theoretical physics community, depending
on the personality of the theorist. The cosmological constant,\cite{weinberg}%
\cite{zeeDirac} that phantom of Einstein's mind, turned out to be real after
all: tiny in the fundamental units of the particle theorists, substantial in
the natural units of the cosmologists, but most definitely not the
mathematicians's zero. The best and the brightest in the theoretical
community missed a golden opportunity much as the opportunity presented to
young Schwinger to calculate the anomalous magnetic moment of the electron.
A brilliant young theorist of our era could have been just as stylish as
Schwinger and calculated, before the observers did their wonderful work, the
cosmological constant to say 3 significant figures, which then turned out to
be in almost exact agreement with the observed value. Alas, that would-be
could-be triumph of string theory did not occur. By the way, before
Schwinger came along, some theorists, equally authoritative as some of our
esteemed colleagues today, also insisted in print that the anomalous
magnetic moment is identically zero.\cite{kosower}

As is well known, the paradox can be easily described. The natural value of $%
\Lambda $ in particle physics is expected by dimensional analysis to be $\mu
^{4}=\mu /(\mu ^{-1})^{3}$ for some relevant mass scale $\mu $ where the
second form of writing $\mu ^{4}$ reminds us that $\Lambda $ is a mass or
energy density$.$ Whether one associates $\mu $ with grand unification,
electroweak symmetry breaking, or the quark confinement transition and
consequently has a value of order $10^{19}$ $Gev,$ $10^{2}$ $Gev,$ and $1$ $%
Gev$ respectively is immaterial. The natural value $\Lambda \sim \mu
^{4}=\mu /(\mu ^{-1})^{3}$ is outrageous even if we take the smallest value
for $\mu .$ We don't even have to put in actual numbers to see that there is
a humongous discrepancy between theoretical expectation and observational
reality. We know the universe is not permeated with a mass density of order
of $1$ $Gev$ on every cube of size $1$ $(Gev)^{-1}.$

The cosmological constant paradox is basically an enormous mismatch between
the units natural to particle physics and natural to cosmology. Measured in
units of $Gev^{4}$ the cosmological constant is so incredibly tiny that
particle physicists have traditionally assumed that it must be
mathematically zero, and have looked in vain for a plausible mechanism to
drive it to zero.

But Nature has a big surprise for us. While theorists racked their brains
trying to come up with a convincing argument that $\Lambda =0,$
observational cosmologists\cite{riess}\cite{perl} steadily refined their
measurements and changed their upper bound to an approximate equality\cite
{neutr} 
\begin{equation}
\Lambda \sim (10^{-3}ev)^{4}!!!
\end{equation}

Thus, theorists once expected the cosmological constant to be ``naturally''
humongous, then insisted that it must be mathematically zero, in spite of
hints\cite{peebles ratra} that it may not be observationally zero. Now
observers have established that it is in fact tiny but not zero.

As is also well known, we should, strictly speaking, refer to the
observation of the cosmological constant as the observation of a hiherto
unknown dark energy since we do not know the equation of state associated
with the observed energy density $(10^{-3}ev)^{4}.$ In this article, we use
the terms ``cosmological constant'' and ``dark energy'' interchangeably.

To make things worse, the energy density $(10^{-3}ev)^{4}$ happens to be the
same order of magnitude as the present matter density of the universe $\rho
_{M}.$ This is sometimes referred as the cosmic coincidence problem. The
cosmological constant $\Lambda $ is, within our traditional understanding, a
parameter in the Lagrangian. On the other hand, since most of the mass
density of the universe resides in the rest mass of baryons, $\rho _{M}(t)$
decreases like $(1/R(t))^{3}$ as the universe expands where $R(t)$ denotes
the scale size of the universe. In the far past, $\rho _{M}$ was much larger
than $\Lambda ,$ and in the far future, it will be much smaller. It just so
happens that in this particular epoch of the universe, when we are around,
that $\rho _{M}\sim \Lambda .$ Or to be less anthropocentric, the epoch when 
$\rho _{M}\sim \Lambda $ happens to be when galaxy formation has been
largely completed. In their desperation, some theorists have even been
driven to invoke anthropic selection\cite{banks}\cite{weinberg2}\cite
{vilenkin}.

In writing this article for these proceedings, I went back and re-read a talk%
\cite{zeeDirac} on the cosmological constant problem I gave just over 20
years ago. To my consternation, and utter humility, I find that many of the
thoughts now current in the theoretical community, including the proton
decay analogy, were already discussed or at least alluded to in that talk.
Incidentally, the talk was given at a conference ``in Honor of P. A. M.
Dirac in his Eightieth Year''.\cite{miami} Unhappily that was the last
occasion I saw Dirac. He died on October 20, 1984 before the proceeding
could be published. As another historical note, J. Schwarz gave a talk on
string theory right after my talk, and since this was before the so-called
string revolution his talk was rather sparsely attended. I remember quite
clearly having lunch with him after his talk and his admonition to me that I
would never understand the cosmological constant without string theory. Be
that as it may, I am glad that John triumphed over his adversaries, namely
all those who left the lecture hall as he started talking about this
at-the-time far-out thing called string theory.

In this article, I would like to highlight two suggestions I made at that
conference. One is the suggestion that the cosmological ``constant'' depends
on the age of the universe. The other is a potentially fruitful analogy with
the story of proton decay. The outline of this article is as follows: after
discussing these two suggestions I list some of the other points I made in
that talk\cite{zeeDirac}, some of which may still prove to be relevant.

\section{ A cosmological constant decaying linearly in time}

In that talk, since the occasion was in honor of Dirac, I was inspired by
Dirac's large number hypothesis and ventured to propose for the cosmological
constant the formula 
\begin{equation}
\Lambda \sim \frac{m_{N}^{3}}{\alpha ^{2}\tau }  \label{propo}
\end{equation}
where $\tau $ denotes the age of the universe. We obtain an oxymoronic
``cosmological constant'' that decays with time. Recall that Dirac, faced
with a retreat from the proposition that large dimensionless numbers should
not appear in physics, suggested that the two large numbers $\alpha
/(Gm_{N}^{2})$ and $m_{N}\tau $ should be equal to each other. I obtained (%
\ref{propo}) in the same spirit, starting with the large number obtained by
dividing the expected value of the cosmological constant by the supposed
upper bound considered reasonable at that time$.$ At present, the proposed
formula would work better numerically if I moved the fine structure constant
squared from the denominator to the numerator: 
\begin{equation}
\Lambda \sim \frac{\alpha ^{2}m_{N}^{3}}{\tau }  \label{formula Cosmo}
\end{equation}
If I take the age of the universe $\tau $ to be $15$ $\times 10^{10}$ years $%
=4.7\times 10^{17}$ seconds which corresponds to an energy of $1.4\times
10^{-33}$ electron volts, the proposed formula (\ref{propo}) gives 
\begin{equation}
\Lambda \sim (3\times 10^{-3}ev)^{4}  \label{alt1}
\end{equation}
Of course, numerically, it is much less sensitive to quote the mass scale of
the physics responsible for the cosmological constant than to quote the
cosmological constant itself. As long as we are playing around we might note
the alternative formulas 
\begin{equation}
\Lambda \sim \frac{m_{e}^{3}}{\tau }  \label{alt2}
\end{equation}
which gives $\Lambda \sim (10^{-4}ev)^{4}$ and 
\begin{equation}
\Lambda \sim \frac{(\alpha _{s}m_{N})^{3}}{\tau }  \label{alt3}
\end{equation}
with $\alpha _{s}$ the strong interaction coupling constant. Needless to
say, this is not serious physics and should be taken as exploratory at best.
I am merely suggesting that Dirac's large number hypothesis appears to lead
to a dark energy density decaying linearly with time.

In the matter dominated era ($\frac{\dot{R}}{R})^{2}\propto \rho _{M}$ and
so the matter density $\rho _{M}$ scales with time like $\rho _{M}\propto
1/t^{2}.$ Thus, a dark energy density decaying linearly with time would not
affect the rather impressive results on primordial nucleosynthesis.

\section{ The proton decay analogy}

The story of the cosmological constant can be summarized as follows: it is a
measurable physical quantity once thought to have the value $\lesssim \infty
,$ then thought to be 0, and finally measured to have the value $\neq 0$ but 
$\lll \infty .$

I asked myself, ``Has there ever been a physical quantity that has gone
through such a dramatic double `reversal of fortune'?''

I came up with the proton decay rate. Recently, I wrote a speculative paper%
\cite{zdark} pointing out this historical precedent and suggesting that it
may shed some light on the resolution of the cosmological constant paradox.
Remarkably, when I wrote that paper I totally forgot that I had already
published it in the proceedings of my 1983 talk\cite{zeeDirac} and referred
only to my field theory textbook.\cite{zee}

Let us go through the story of proton decay. To make my point I will take
some liberty with history. Suppose that in 1953 some theorists were to
calculate the rate $\Gamma $ for protons to decay in the natural mode $%
p\rightarrow e^{+}+\pi ^{0}.$ The interaction of the pion with the proton
and the neutron was known to be described by a term like $g\pi \overline{n}p$
in the Lagrangian with $g$ a dimensionless coupling of order 1. These
theorists would naturally construct a Lagrangian out of the available
fields, namely the proton field $p,$ the electron field $e,$ and the pion
field $\pi ,$ and thus write down something like $f\pi \overline{e}p$ with
some constant $f.$ Note that $\pi \overline{e}p$ has mass dimension 4 and
hence $f$ is dimensionless just like $g$. Since $\pi \overline{e}p$ violates
isospin invariance, the theorists would expect $f$ to be suppressed relative
to $g$ by some measure of isospin breaking, say the fine structure constant $%
\alpha .$ The natural value for $\Gamma $ would then come out to be many
many orders of magnitude larger than the experimental upper bound on $\Gamma
.$ The theorists would then set $\Gamma =0$ and cast about for an
explanation. After an enormous struggle, the theorists were unable to come
up with a compelling explanation and this failure became known as the proton
decay rate paradox.

Eventually, someone with great authority and prestige in the community,
namely Wigner, decreed the law of baryon number conservation. Surely, even
in the unthinkably primitive days of 1953 this would have been recognized as
a pronouncement and not as an explanation. (The pronouncement could be
dressed up formally by imposing a $U(1)$ transformation under which $%
p\rightarrow e^{i\theta }p$ while $e$ and $\pi $ do not change and requiring
that the Lagrangian remains invariant.) But there would have been no deep
understanding of this astonishing discrepancy between theoretical
expectation and experimental upper bound.

Indeed, imagine an alternative history in which, while other important
particle physics experiments were being performed in 1957, some intrepid
experimentalist, ignoring conventional theoretical wisdom, actually went out
and measured the proton decay rate to be some tiny but non-zero value. The
proton decay rate paradox would have deepened, much as how the cosmological
constant paradox deepened with the discovery of a tiny $\Lambda .$

Let us now review how the proton decay rate paradox was resolved
historically. The first remark is that the eventual explanation did not
emerge within the orthodox theory fashionable in 1957, nor did it come from
an understanding of some kind of mechanism causing protons to decay, but
rather it came totally from ``left field'', from a study of baryon
spectroscopy, which led to the notion of quarks. The correct degrees of
freedom are not given by the proton and pion fields $p$ and $\pi ,$ but by
the quark fields $q.$ The effective Lagrangian ${\cal L}$ is to be
constructed out of quark $q$ and lepton $l$ fields and must satisfy the
symmetries that we know. Three quarks disappear, so we write down
schematically $qqq,$ but three spinors do not a Lorentz scalar make. We have
to include a lepton field and write $qqql.$ Since four fermion fields are
involved, these terms have mass dimension 6 and so in ${\cal L}$ they have
to appear as $\frac{1}{M^{2}}qqql$ with some mass $M,$ corresponding to the
mass scale of the physics responsible for proton decay. Thus, the
probability of proton decay is proportional to ($\frac{1}{M^{2}})^{2}=\frac{1%
}{M^{4}}.$ By dimensional reasoning, we obtain the proton decay rate $\Gamma
\sim (\frac{m_{p}}{M})^{4}m_{p}.$ The absurdly small value of $\Gamma $ is
then naturally explained by the fourth power of the small number $(\frac{
m_{p}}{M})$ for $M$ big enough. No mystery left!

Note that in principle all of this could be done as soon as Gell-Mann
introduced the notion of quarks in 1964, long before anybody even dreamed of
a grand unified theory with proton decay.

As long we are discussing revisionist, but possible, history, we can imagine
some brilliant theorist in another civilization far far away puzzling over
the proton decay rate paradox eventually realizing that the key to
explaining an absurdly small number is to promote the dimension of the
effective Lagrangian merely from 4 to 6. In hindsight, we can say that the
extremely long lifetime of the proton could have pointed to the existence of
quarks.

I would like to raise the question whether the cosmological constant paradox
might not be solved in the same way. Perhaps the gravitational field $g_{\mu
\nu }$ is the analog of the proton and pion field $p$ and $\pi .$ The high
energy and more fundamental degrees of freedom in the gravitational field
may not be the metric $g_{\mu \nu },$ but some mysterious analog of the
quark field $q.$ This may emerge as a construct in string or M theory, or it
could be something else completely. In the history of the proton decay
paradox as recounted by me, there is an additional twist, namely that the
degree of freedom $q$ is confined and not physical. Before the advent of
quantum chromodynamics, theorists could only write $p\sim qqq,$ without any
clear idea about what the symbol $\sim $ might mean. We are in a similar
position here: the metric $g_{\mu \nu }$ might be a composite object, but I
certainly do not know what it is a composite of, and the objects of which $%
g_{\mu \nu }$ is a composite may also be as observable or as unobservable as
the quarks. Just as we did not know what the symbol $\sim $ might mean in $%
p\sim qqq,$ we may have no idea of even what the word ``composite'' might
mean in the case of $g_{\mu \nu }.$

The cosmological term $\Lambda \sqrt{g}$ in the Lagrangian has mass
dimension 0 and we somehow have to promote 0 to a higher number. One
difficulty with this view is of course how we could possibly promote the
dimension of the cosmological term without at the same time changing the
mass dimension of the Einstein-Hilbert term $\frac{1}{G}\sqrt{g}R.$ Our
historical analogy may again be helpful: the 1953 view that the pion nucleon
coupling term has dimension $4$ turns out to be correct. While the dimension
4 term $\pi \overline{e}p$ was replaced by the dimension 6 term $qqql$ the
dimension 4 term $\pi \overline{n}p$ was replaced by dimension 4 terms of
the form $\overline{q}Aq$ with $A$ a gluon potential. The dimension of one
of the two terms gets promoted while the dimension of the other term remains
the same. So it is entirely conceivable to me that the cosmological constant
term could end up with a higher dimension while the Einstein-Hilbert term
either remains dimension 4 or is replaced by dimension 4 terms. Thus,
suppose the cosmological constant term actually has dimension $p>0$ so that
it is given in the Lagrangian by a term of the form $\frac{1}{M^{p-4}}{\cal O%
}$ with $M$ some mass scale characteristic of the deeper structure of the
graviton, perhaps the same as the Planck mass, perhaps not. The observed
cosmological constant would then be given by $\Lambda \sim \frac{1}{M^{p-4}}<%
{\cal O}>=(\frac{m}{M})^{p}M^{4},$ where the expectation value of the
operator ${\cal O}$ in the physical universe $<{\cal O}>=m^{p}$ is set by
physics at some low energy scale $m.$ With $m$ small enough, and or $p$ big
enough, we could easily get the suppression factor we want.

\section{Nature of gravity}

One difficulty with this analogy may be that it relies implicitly on our
understanding of quantum field theory as low energy effective field
theories. Banks, Fishler, and others have emphasized to me that this
customary understanding may not apply to gravity. Our formulation of field
theory relies on that marvellous invention known as the Fourier transform $%
e^{ipx}$ according to which low momentum and energy corresponds to long
distances, so that microscopic details do not enter the effective theory.

However, while this appears to govern all particles it does not appear to
work for black holes. According to the celebrated relation $R=GM$ the more
massive a black hole the larger it is. The size of a black hole is not
governed by its Compton wavelength! Evidently, this strange result follows
from the fact that gravity, in contrast to the other fundamental
interactions, has a dimensionful coupling constant. The relation $R=GM$
follows from the dimension of $G$ and the supposition that black holes
should not exist as we turn off gravity (supplemented by a certain notion of
simplicity.) We could imagine (although we do not know how to actually do
it) modifying Einstein's theory so that this relation holds for astronomical
black holes but not for microscopic black holes that we should include in
our effective field theories. Perhaps $G$ depends on $M$ in such a way that
we could recover the deBroglie-Compton relation that $R\sim M^{-1}.$ Or
perhaps in the standard form of the Schwarzschild solution $g_{00}$ is
modified to $(1-\frac{GM}{f(r)})$ so that $f(r)\rightarrow r$ for large $r$
but behaves differently for small $r,$ so that the dependence of $R$ on $M$
is changed. These of course are just ``empty'' remarks at this point.

Another point Banks\cite{banksP} has impressed upon me is a possible
connection between gravity and quantum measurement theory. In the usual
formulation, a classical measuring apparatus, classical because it could be
made very large and massive, is needed to study a quantum system. All very
well without gravity! But with gravity, the massive classical apparatus
would exert an unacceptable gravitational effect on the system we are trying
to measure. To minimize this perturbation we are obliged to move the
classical apparatus to infinity, which according to Banks gives a whiff of
`t Hooft's holographic principle. I find this very disturbing, to say the
least.

While invoking the proton decay analogy I do not necessarily insist that the
graviton is composite, although that is certainly a possibility. The
graviton could be part of a larger structure.\cite{gravitipole} Inspired by
the almost exact correspondence between Einstein's post-Newtonian equations
of gravity and Maxwell's equations of motion I once proposed the gravitipole
in analogy with Dirac's magnetic monopole. After Dirac there was
considerable debate on how a field theory of magnetic monopoles may be
formulated. Eventually, 't Hooft and Polyakov showed that the magnetic
monopole exists as an extended solution in certain non-abelian gauge
theories. Most theorists now believe that electromagnetism is merely a piece
of a grand unified theory and that magnetic monopoles exist. Might it not
turn out that Einstein's theory is but a piece of a bigger theory and that
gravitipoles exist? In grand unified theory the electromagnetic field is a
component of a multiplet. Could it be that some generalized gravitational
field ${\cal G}_{\mu \nu }^{\alpha }$ also somehow carries an internal index 
$\alpha $ and that the gravitational field we observe is just a component of
a multiplet? Throwing caution to the wind, I also asked in\cite{gravitipole}
if the gravitipole and the graviton might not form a representation under
some dual group just as the magnetic monopole and the photon form a triplet
under the dual group of Montonen and Olive\cite{mo}.

In this connection, I have always been quite taken by 't Hooft's elegant
expression for Maxwell's electromagnetic field 
\[
{\cal F}_{\mu \nu }=\frac{F_{\mu \nu }^{a}\varphi ^{a}}{|\varphi |}-\frac{%
\varepsilon ^{abc}\varphi ^{a}(D_{\mu }\varphi )^{b}(D_{\nu }\varphi )^{c}}{%
|\varphi |^{3}} 
\]
in terms of the Yang-Mills field. I wonder from time to time if there might
not be an analogous expression, giving Einstein's gravitational field $%
g_{\mu \nu }$ in terms of ${\cal G}_{\mu \nu }^{\alpha }.$

\smallskip

\section{ Miscellaneous thoughts}

There is of course no lack of speculative thoughts\cite{sundrum} about the
cosmological constant in the literature. Let me take this opportunity to
list some of the other points I mentioned in my talk\cite{zeeDirac}.

(1) The fact that gravity is not renormalizable is not really relevant as
long as we wish to study only gravitational physics at low energy and long
distances. We have an effective low energy theory consisting of an expansion
in powers of ($1/M_{Pl})^{2}$ and with radiative corrections cut off at some
unknown mass scale $M_{C}.$ The cosmological constant and the
Einstein-Hilbert action correspond to the first two terms.

(2) The graviton knows about anything carrying energy and momentum,
including an apparently innocuous constant shift in the Lagrangian density $%
{\cal L}\rightarrow {\cal L}-\Lambda $. One possible way of nullifying the
physical consequence of the shift ${\cal L}\rightarrow {\cal L}-\Lambda $ is
to postulate that $g=\det g_{\mu \nu }$ is not a dynamical variable. This
was proposed\cite{w+z}\cite{FWg=1} as an explanation of why $\Lambda $ is
mathematically zero. But with $\Lambda $ now known to be tiny but non-zero
this avenue seems to me less promising. Historically, it is amusing to note
that Einstein did his first calculation\cite{einstein} imposing the gauge
choice $g=1.$ This approach has been explored by S. Weinberg\cite{weinberg}
and others.

(3) Besides the proton decay analogy mentioned above, I also pointed out 2
other possibly relevant analogies. I mention one of them here. In the days
when the strong interaction was described by the Yukawa theory, it was
mysterious why the strong interaction is parity and charge conjugation
invariant and conserves strangeness. One can perfectly well introduce terms
such as $a\overline{\Sigma }p$ and $m_{N}\overline{p}(1+b\gamma _{5})p$
(here $\Sigma $ denotes the sigma hyperon) and so forth into the Lagrangian.
The ``natural'' expectation would be that the dimensionless parameters $a$
and $b$ are of order unity, while experimentally they are known to be very
small. Even if we had a reason to exclude terms like these they would be
induced by the weak interaction. At that time physicists were at a loss to
explain why these terms are not there. The solution did not come until the
formulation of quantum chromodynamics. Gauge invariance allows the offending
terms to be simply transformed away. As forcefully emphasized by Weinberg,
this represents one of the virtues of quantum chromodynamics.

This example may be telling us that perhaps Einstein's theory is to the true
theory of gravity as Yukawa's theory is to quantum chromodynamics.

(4) I discussed the possibility of promoting $\Lambda $ into a dynamical
field $\Lambda (x),$ which I mentioned is tantamount to introducing a scalar
field $\varphi (x).$ Somehow $\varphi (x)$ would move dynamically to counter
any ``bare'' $\Lambda .$ To do this, the scalar field must know about $%
\Lambda ,$ and thus must be coupled to the curvature in various ways. I
believe that some versions of this kind of approach is captured in the
contemporary literature in the notion of ``tracker fields''.

(5) The cosmological constant is a property of the particular ground state
of the Universe at a particular instant. For example, in induced gravity\cite
{adler}\cite{zeeinduced} $\Lambda $ is given by $<0|T|0>$ and thus depends
on the ground state $|0>$ (here $T$ denotes the trace of the stress energy
tensor of course.) However, the cosmological evolution $|0>$ depends on $%
\Lambda ,$ schematically, $\frac{\partial }{\partial t}|0>=F[\Lambda ]$ for
some functional $F,$ so that $\frac{\partial \Lambda }{\partial t}$ is a
functional of $\Lambda .$ It is thus conceivable that the cosmological
``constant'' might relax to zero or near zero in the present universe. This
approach was pursued by Adler, Myhrvold, Mottola, and others without
reaching a definitive conclusion.

(6) There is always the possibility\cite{feynmanQFT} that the quantum field
theory formulation of the ``vacuum'' is incorrect or inconvenient. In modern
times, this is perhaps represented by the thoughts behind the holographic
principle as outlined above by Banks. But I am rather disinclined not to
trust quantum field theory; at the very least, quantum field theory has
already survived two near-death experiences.

Perhaps we do not know as much about the graviton as we think we do.
Possibly one concrete sign of this is the amazing discovery by Bern\cite
{bern} and others that the Einstein-Hilbert action may factorize order by
order in the gravitational field into what one might think of as a product
of two Yang-Mills actions.

\smallskip

{\bf {\large Acknowledgments\medskip }}

This work was supported in part by the National Science Foundation under
grant number PHY 99-07949. \medskip I am also grateful to Professor P. Hwang
for his warm hospitality. I appreciate discussions with Tom Banks in writing
up this talk for the Proceedings.

\end{document}